# Electronic correlations determine the phase stability of iron up to the melting temperature

I. Leonov,[1] A. I. Poteryaev,[2,3] Yu. N. Gornostyrev,[2,3]

A. I. Lichtenstein,[4] M. I. Katsnelson,[5,6] V. I. Anisimov,[2,6] and D. Vollhardt[1]

[1]Theoretical Physics III, Center for Electronic Correlations and Magnetism, Institute of Physics, University of Augsburg, 86135 Augsburg, Germany

[2]Institute of Metal Physics, 620219 Yekaterinburg GSP-170, Russia

[3]Institute of Quantum Materials Science, 620107 Yekaterinburg, Russia

[4]Institute of Theoretical Physics, University of Hamburg, 20355 Hamburg, Germany

[5]Radboud University Nijmegen, Institute for Molecules and Materials, 6525 AJ Nijmegen, Netherlands

[6]Ural Federal University, 620990 Yekaterinburg, Russia

Corresponding author: Ivan.Leonov@physik.uni-augsburg.de

**We present theoretical results on the high-temperature phase stability and phonon spectra of paramagnetic bcc iron which explicitly take into account many-body effects. Several peculiarities, including a pronounced softening of the [110] transverse ($T_1$) mode and a dynamical instability of the bcc lattice in harmonic approximation are identified. We relate these features to the α-to-γ and γ-to-δ phase transformations in iron. The high-temperature bcc phase is found to be highly anharmonic and appears to be stabilized by the lattice entropy.**

The extraordinary magnetic and metallurgical properties of elemental iron (Fe) have attracted the attention of mankind for several thousand years already. In particular, iron is the main ingredient of steel which is one of the most important construction materials of modern civilization.[1] In this respect we still live in the "iron age". Iron is also of primary interest for geophysics as the main constituent of the Earth's core.[2,3,4,5,6,7] Furthermore, iron is perhaps the most famous example of allotropy in metals. Indeed, iron is a very complicated material where magnetic order[8,9,10,11,12,13,14,15,16] and correlation effects[17,18,19,20,21,22] play an important role. For these reasons the electronic and lattice properties of iron are still far from understood.

Iron has at least four allotropes: the α, γ, and δ phases at low, and the ε phase at high pressure, respectively. It is particularly remarkable that the low-temperature structure of iron, the ferromagnetic α phase, is not close-packed but body-centered cubic (bcc). This already indicates that iron has highly unusual features. In fact, among the elements iron is characterized by the unique property of having *two* stable bcc phases: the aforementioned α phase, which is stable below 1185 K, and the δ phase which is stable above 1670 K up to the melting curve. Sandwiched in between those two phases is the γ phase ("austenite") which has a closed-packed face-centered cubic (fcc) structure. It is understood that magnetism (iron is ferromagnetic below the Curie temperature $T_C$ ~ 1043 K) stabilizes the low-temperature α phase by compensating for the too low cohesive energy. Indeed, the low-temperature bcc phase does not exist for nonmagnetic Ru and Os, which lie below Fe in the periodic table. Moreover, iron makes a transition to a hexagonal closed-packed (hcp) structure (ε phase) once ferromagnetism is suppressed by pressure. Clearly it is the interplay between magnetism of the electrons and the lattice degrees of freedom which determines the complexity of the low-temperature phase diagram of iron.

At high temperatures, in the absence of ferromagnetic long-range order, the situation is even less clear. In general, polymorphic metals which have both bcc and closed-packed structures (e.g., Fe, Ca, Sr, Ti, Zr, Ce, Pu) are known to melt from the bcc phase. As argued by Zener,[23] this can be caused by a somewhat lower Debye temperature in a bcc lattice. Namely, the bcc phase has a low-energy $T_1$ phonon [transverse (110)] branch, whose frequencies vanish in the case of pair-wise, nearest-neighbor interatomic interactions. Such a soft-mode behavior is associated with a large amplitude of vibrations. It leads to a high lattice entropy, which lowers the free energy and thereby stabilizes the bcc lattice. This behavior is clearly revealed in prototypical systems with metallic chemical bonds, such as alkali and alkaline-earth metals,[24] as well as in the high temperature bcc phase of the group 3 (Sc, Y, La) and 4 (Ti, Zr, Hf) metals.[25] By contrast, the phonon modes in bcc transition metals like Cr, Mo, and W, where the valence d-band is almost half-filled and the chemical bond is covalent, are hard. Moreover, these metals are not polymorphic.[26]

It is quite surprising to note that in the case of iron the experimental information about electronic and dynamical lattice properties at high temperatures is still very limited. In fact, to the best of our knowledge results on the lattice dynamical properties of δ iron have not available. In particular, soft-mode behavior of the phonon spectra has neither been reported at the α-to-ε phase transition, nor in the γ phase.[27,28,29] Thus the lattice dynamics, in particular its evolution at high temperatures and near the α-to-γ and γ-to-δ phase transitions in iron, remain unexplored.

An overall explanation of the electronic and lattice properties of iron at high temperatures requires a formalism which can treat both paramagnetic and magnetically ordered states on equal footing and can describe local moments above $T_C$. Therefore conventional band-structure techniques cannot be used to describe the α-to-γ phase transition in *paramagnetic* iron, although they provide a good quantitative description of the properties of ferromagnetic α iron.[9,10,11,12,13] By contrast, the structural transition actually occurs at a temperature $T_{struc}$ which lies about 150 K *above* $T_C$. Moreover, these techniques find the bcc phase to be elastically unstable in the absence of magnetization. Today we know[20,21] that an improved description of these transitions can only by obtained when the existence of local moments above $T_C$ is taken into account.

To clarify these fundamental questions we performed microscopic calculations of the structural phase stability and the phonon spectra of iron within the so-called LDA+DMFT approach.[30,31,32] It merges *ab initio* techniques for the calculation of band structures, such as the local density approximation (LDA) or the generalized gradient approximation (GGA), with dynamical mean-field theory (DMFT)[33,34,35] – a state-of-the-art computational scheme for the calculation of the electronic structure of strongly correlated materials. This combined approach allows one to compute the properties of correlated materials in both their paramagnetic or magnetically ordered states. Several applications of this scheme have provided very good quantitative descriptions of the electronic, magnetic, and structural properties of iron.[17,18,19,20,21,22] In particular, the LDA+DMFT method implemented with plane-wave pseudopotentials[36] has recently been employed to compute the equilibrium crystal structure and phase stability of iron near the α-to-γ phase transition.[20,21] Electronic correlations were found to be essential for the explanation of the electronic and structural properties of paramagnetic iron. For example, the α-to-γ phase transition was found to occur at ~ 1.3 $T_C$, i.e., well above the magnetic transition, in agreement with experiment. Moreover, the calculated phonon spectra of iron near the α-to-γ phase transition were also shown to be in good agreement with experiment.

**Results**

We computed the temperature evolution of the lattice dynamical properties of iron near the α-to-γ and γ-to-δ phase transformations within LDA+DMFT for Coulomb interaction parameters $U = 1.8$ eV and $J = 0.9$ eV as obtained by earlier theoretical and experimental estimations.[37,38] Our calculations of the Curie temperature in bcc Fe yield $T_C = 1650$ K, which is in reasonable agreement with the experimental value 1043 K given the local nature of the DMFT approach[32,33,34,35] and the approximate density-density form of the Coulomb interaction.[22] To extract the phonon frequencies, the dynamical matrix was evaluated using the method of frozen phonons[39] implemented within LDA+DMFT. By introducing small displacements of atoms from their equilibrium positions, we determined the phonon frequencies at given **q**-points. The phonon dispersion curves were obtained by using a standard Born–von Kármán (BvK) force constant model with interactions expanded up to the fifth-nearest neighbor. In the following all results will be presented as a function of the reduced temperature $T/T_C$.

Our results for the phonon dispersion curves of iron near the magnetic phase transition temperature $T_C$ are shown in Fig. 1. The calculated phonon dispersions exhibit the typical behavior of a bcc metal. The phonon frequencies are overall positive, *both in the ferromagnetic and paramagnetic phase*. This implies the stability of the bcc lattice structure of iron in accordance with experiment. We also note the remarkably good quantitative agreement between the calculated phonon dispersions of bcc iron and the experimental data[28] taken at T ~ $T_C$ (a detailed comparison at 1.2 $T_C$ was reported earlier[21]). In the given temperature range the calculated phonon dispersions show only a rather weak temperature-dependence. However, there is an anomaly in the transverse (110) acoustic mode ($T_1$ mode) along the Γ-N branch near the bcc-fcc phase transition temperature at T ~ 1.2 $T_C$. This can be ascribed to a dynamical precursor effect of the bcc-to-fcc phase transition which occurs at $T_{struct}$ ~ 1.3 $T_C$ in our calculations.[20]

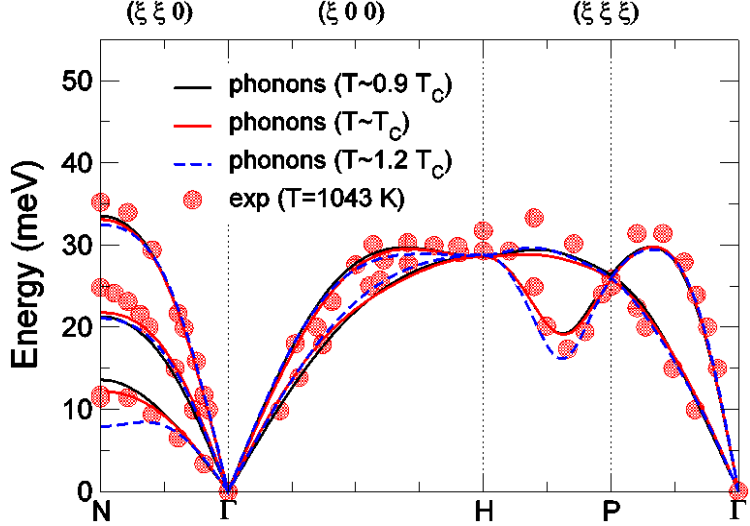

**Figure 1:** Calculated phonon dispersion curves for bcc iron near the Curie temperature $T_C$. The results are compared with neutron inelastic scattering measurements at 1043 K.

Next we discuss the lattice dynamics of paramagnetic bcc iron in more detail. Our results (Fig. 2), obtained in the harmonic approximation, clearly indicate that at T ~ 1.4 $T_C$, i.e., above the α-to-γ phase transition temperature, the bcc phase is dynamically unstable. The origin of the instability lies in the $T_1$ mode in the Γ-N direction whose frequency becomes imaginary near the N-point. At the same time other phonon branches remain stable and are weakly temperature dependent near the transition point. At even higher temperatures, T ~ 1.8 $T_C$, in the δ phase, the bcc lattice structure is dynamically unstable in the whole Γ-N branch, again due to the $T_1$ mode, with an additional anomaly near the (2/3,2/3,2/3) point. The latter anomaly may be interpreted as a precursor effect similar to the one occurring at the β-ω transition in Ti and Zr.[25,40] Therefore our calculations reveal the crucial importance of the $T_1$ phonon mode for the explanation of the structural phase stability of *paramagnetic* bcc iron, in agreement with the Zener's hypothesis.[23] Anharmonic lattice effects must be included to obtain a complete description of the lattice dynamics of δ iron. This situation is quite similar to that in the ε phase of Pu, which has a bcc structure and which is dynamically unstable in the harmonic approximation.[41] It also seems to be the case in other bcc transition metals, such as Sc, Y, La, Ti, Zr, Hf, and Th.[25,42] However, in the δ phase of Fe and the ε phase of Pu the situation is very special since, both, anharmonic lattice and electron-electron correlation effects are highly relevant, making the problem particularly difficult to investigate.

To illustrate the importance of anharmonic effects in bcc iron, we calculated the total energy profiles for the longitudinal (*L*) and the two transverse ($T_1$ and $T_2$) modes at the N-point. The results are shown in Fig. 3, as a function of the corresponding atomic displacement for three different temperatures. We compare our results with those obtained for the lowest phonon, the transverse (111) mode, of the fcc phase. It is seen that the two bcc phonon modes *L* and $T_2$ are rigid and do not change across the α-to-γ phase transition. Similarly, the fcc (111) phonon mode is stable and is almost temperature independent. By contrast, the $T_1$ mode is strongly affected by temperature: while it is stable at T ~ 1.2 $T_C$ < $T_{struc}$, i.e., below the calculated α-to-γ phase transition temperature, a shallow double-well potential develops at T ~ 1.4 $T_C$ > $T_{struc}$, which is responsible for the lattice instability. Our estimate of the phonon free energy, which is based on the method described in Ref. 43, shows that the bcc lattice is favored over the fcc lattice since the corresponding phonon free energy is lower by ~ 0.007 eV/atom. This, however, is not sufficient to stabilize the bcc phase, since the *electronic* energy of the fcc phase is substantially (by ~ 0.03 eV/atom) lower than that of the bcc phase (see Fig. 3b).

Therefore the fcc lattice structure is the stable one. At even higher temperatures, T ~ 1.8 $T_C$, the $T_1$ mode becomes strongly anharmonic, with a pronounced double-well structure. It results in a high lattice entropy which lowers the phonon free energy and hence favors the bcc lattice structure over the fcc lattice by ~ 0.055 eV/atom. This is sufficient to overcome the electronic energy loss of ~ 0.04 eV/atom (see Fig. 3c) and to stabilize the bcc phase at high temperatures. On the basis of our results we therefore conclude that the high-temperature bcc (δ) phase of iron is stabilized by the lattice entropy, which gradually increases upon heating due to the increasingly anharmonic behavior of the $T_1$ phonon mode.

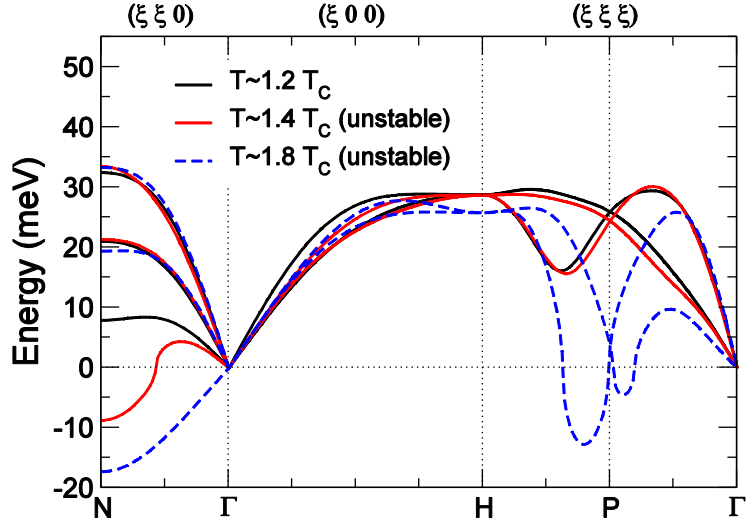

**Figure 2:** Calculated phonon dispersions of paramagnetic bcc iron near the α-to-γ and γ-to-δ phase transitions for different temperatures.

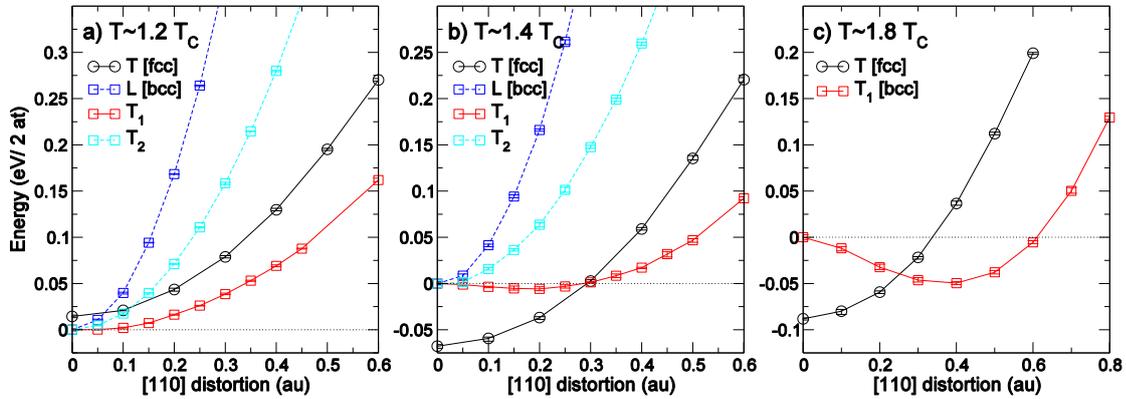

**Figure 3:** Calculated total energy of paramagnetic iron as a function of displacement (in au) for the longitudinal (*L*) and the two transverse ($T_1$ and $T_2$) phonon modes at the N point for the bcc lattice, and the transverse (111) mode at the L-point for the fcc lattice, respectively.

Our results for the temperature evolution of the lattice dynamics offer a new perspective on the microscopic origin of polymorphism of iron. We find that the $T_1$ transverse (110) phonon mode is responsible for the α-to-γ and γ-to-δ allotropic transformations in iron. Furthermore we predict that this mode continuously softens upon temperature increase – a feature which is typical for simple nonmagnetic metals.[25,42] For T > $T_{struc}$ ~ 1.3 $T_C$ the bcc lattice becomes dynamically unstable near the N-point and collapses at ~ 1.8 $T_C$, at least in the harmonic approximation. By contrast, at temperatures T < $T_{struct}$ the phonon modes of bcc iron are relatively rigid, and resemble those of non-polymorphic bcc metals such as Cr, Mo, W. The phonon spectra of the fcc (γ) iron show only a rather weak temperature dependence.[21] For the bcc phase the phonon spectra are strongly affected by temperature, i.e., the phonon frequencies decrease as the temperature is raised. We found that the $T_1$ phonon mode softens at the fcc-to-bcc phase transition, which indicates the importance of anharmonic effects for a correct description of the high-temperature δ phase of iron. Obviously, lattice anharmonicity and electron-electron correlations have to be taken into account simultaneously. This is a great challenge for future research.

**Methods**

We employed the LDA+DMFT scheme[30,31,32] to compute lattice dynamical properties of iron in the temperature range covering the α-to-γ and γ-to-δ phase transitions. This state-of-the-art method combines *ab initio* density-functional-based techniques, e.g., the local density approximation (LDA) or general gradient approximation (GGA), with dynamical-mean field theory (DMFT) for correlated electrons systems, which describes the quantum dynamics of the many-electron problem exactly but neglects non-local effects. The LDA+DMFT approach allows one to include the effect of electronic correlations on the electronic band structure and phonon properties of correlated materials. To include the electronic correlations we derived an effective low-energy Hamiltonian for the valence Fe *sd* orbitals. We employed the non-magnetic GGA together with plane-wave pseudo-potentials to construct a basis of atomic centered symmetry constrained Wannier functions for the Fe *sd* orbitals.[36] The multi-orbital Hubbard Hamiltonian obtained thereby, which includes the Coulomb repulsion $U$ and Hund's exchange interaction $J$ for the 3d Fe orbitals, is solved by DMFT using quantum Monte Carlo (QMC) simulations.[44] To determine the phase stability and lattice dynamics, we calculated the total energy and evaluated the corresponding phonon frequencies using the frozen-phonons method.[39] The interaction parameters $U = 1.8$ eV and $J = 0.9$ eV are assumed to remain constant during the structural transformations. Further technical details about the method used can be found in Ref. 36.

**Acknowledgements**

We thank Igor Abrikosov, Michael Leitner, and Winfried Petry for helpful discussions. This work was partially supported by the Deutsche Forschungsgemeinschaft through TRR 80 (I.L.) and FOR 1346 (A.I.L., D.V.), by the Russian Foundation for Basic Research through Project Nos. 13-02-00050 (V.I.A.) and 13-03-00641 (A.I.P.). M.I.K. acknowledges a support by the Nederlandse Organisatie voor Wetenschappelijk Onderzoek (NWO) via Spinoza Prize.

**Author contributions**

I.L., V.I.A. and D.V. conceived and supervised the project. I.L. performed the computations. All authors contributed to the interpretation of the data and to the writing of the manuscript.

**Competing financial interests**

The authors declare no competing financial interests.